\documentclass[12pt]{iopart}

\usepackage{graphicx}
\usepackage{color}
\begin{document}

\title{Solar neutrino interactions with liquid scintillators  used for 
double beta-decay experiments}

\author[first]{Hiroyasu Ejiri$^1$ and Kai Zuber$^2$}

\address{1. Research Center for Nuclear Physics, 
Osaka University, Osaka 567-0047, Japan \\
2. Institute for Nuclear and Particle Physics, TU Dresden, 01069, Dresden, Germany
}
\ead{ejiri@rcnp.osaka-u.ac.jp, zuber@physik.tu-dresden.de}

\vspace{10pt}
\begin{indented}
\item[] 
\end{indented}

\begin{abstract}
Solar neutrinos interact within double-beta-decay (DBD) detectors and hence will contribute to backgrounds (BG) for DBD experiments. 
Background contributions due to solar neutrinos are evaluated for their interactions with atomic electrons 
and nuclei in liquid scintillation detectors used for DBD experiments. They are shown to be 
serious backgrounds for high-sensitivity DBD experiments to search for the Majorana neutrino masses in the inverted and normal hierarchy regions.    \\

\end{abstract}

\section{Introduction}
Neutrino-less double beta decays (0$\nu \beta \beta $) are crucial for studies of the neutrino ($\nu $) properties and especially the Majorana mass character of the neutrino. With the given current knowledge from neutrino oscillation experiments, the typical mass regions to be 
explored are about 45-15 meV and 4-1.5  meV in cases of the inverted hierarchy (IH) and normal hierarchy (NH) mass spectra, respectively. 
Since the $0\nu \beta \beta $ half-lives expected for these regions are well above $10^{26-27}$ years, the 0$\nu \beta \beta $ 
rates are so low that large scale detectors have to be used.
The amount of double-beta-decay (DBD) isotopes for such experiments are  multi tons and multi 100 tons for IH and NH masses, also 
depending on the nuclear matrix element and the phase space volume.
Background rates, thus, have to be necessarily as low as an order of  or less than 1 per year (y) per ton (t) 
in the region of interest (ROI) at the $Q$-value 
in the sum energy spectrum of the two $\beta $ rays. DBD studies and $\nu $ masses are discussed in recent reviews and their references \cite{eji05,zub98,rod11,ver12}. 

The solar-$\nu$s are omnipresent and cannot be shielded, and thus their 
charged current (CC) and neutral current (NC) interactions are potential background (BG) sources 
for high sensitivity DBD experiments \cite{eji14,bar11} and references therein. 
Actually, it has been shown that solar-$\nu $ CC interactions with DBD isotopes 
like $^{100}$Mo, $^{116}$Cd and $^{150}$Nd can be used for real-time studies of the low-energy solar-$\nu$s 
\cite{eji00,zub03,zub12}. 

The solar-$\nu$ NC and CC interactions with atomic electrons ( $\nu $-e scattering) and nuclei in DBD detector are considered. 
Fig.1 shows the interaction schemes for the CC interactions with the atomic electron and the nucleus.
The scattered electron, the electrons (inverse $\beta $ ray) from the CC interaction, 
the $\gamma $ rays following NC and CC interactions, and delayed $\beta $ rays from 
the intermediate nucleus B contribute to the BGs if they deposit their energy at the ROI of the 0$\nu \beta \beta $ $Q$ value. 

\begin{figure}[htb]
\caption{Solar-$\nu$ CC interaction schemes. Left hand side: Solar-$\nu $ CC interaction 
with an atomic electron in a DBD detector. 
Right hand side: Solar-$\nu $ CC interaction with a nucleus A  in a detector. e: scattered electron. W: CC weak boson. p: proton.
n: neutron. B: residual nucleus. $^\ast$ indicates a possible excited state in B.
\label{fig:strengths}}
\begin{center}
\includegraphics[width=0.5\textwidth]{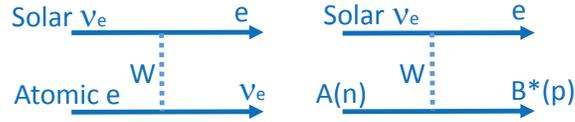}
\end{center}
\end{figure}

Large scale DBD experiments are under progress and/or in discussion 
to search for the Majorana $\nu $ masses in the IH and NH regions. 
One discussed option is the usage of large liquid-scintillator detectors loaded 
with the DBD isotope of interest at the level of some 10$^{-2}$ of the scintillator mass itself or even larger. 
KamLAND-Zen uses a k-ton liquid scintillator 
with multi-100 kg $^{136}$Xe \cite{kam12}, and SNO+ plans to study $^{130}$Te  by means of a 
0.8 k ton liquid scintillation detector loaded with 0.3 \% natural Te ($^n$Te) \cite{sno14}. Both consider upgrades in
the amount of loaded DBB isotope and other new experiments might follow this idea.
Liquid scintillators have also been 
discussed for possible $\beta \beta $ experiments with $^{100}$Mo for the MOON experiment 
\cite{eji00,geh10} and $^{150}$Nd \cite{sno11}
at the beginning of SNO+.
The total volume of the detector amounts to k-ton, and larger detectors like JUNO with 20 k-tons are 
already in preparation and M-ton scales might be necessary for NH mass studies. 
Given these sizes of detectors, BGs due to solar $\nu $ interactions 
with liquid scintillators have to be considered as well, as already 300 ton scintillation detectors like
Borexino operate as solar-$\nu$ experiments. 

In the previous papers, 
we evaluated BG contributions of the solar-$\nu$
 interactions with DBD nuclei \cite{eji14} and atomic electrons of DBD isotopes \cite{bar11}. 
 In the latter case an average number of about 1.5 x 10$^{-4}$ counts/keV per year ton (yt) for all relevant DBD 
isotopes is expected due to the $^8$B-$\nu$ interactions. 
The solar-$\nu $ interactions with DBD isotopes were shown to be serious for DBD experiments to search for the 
 IH and NH mass regions, depending on the nuclear matrix element and the energy resolution of the detector at the $Q$-value \cite{eji14}.
 
The present paper aims to discuss contributions of the solar-$\nu$ interactions with atomic electrons and nuclei involved in 
large-scale liquid scintillators.  This involves the DBD isotope and the scintillator material. Actually the solar-$\nu$ 
BG may set an upper limit on the DBD detector sensitivity for the $0\nu \beta \beta $ half-life to be measured and thus a lower limit of the sensitivity for the 
Majorana neutrino mass to be detected.\\

\section{ Solar-$\nu$ interactions with scintillators}
Let us first evaluate event rates of the solar-$\nu $ interactions with atomic electrons and nuclei in
liquid scintillators to 
compare them with 0$\nu \beta \beta $ rates of current $\beta \beta $ experiments for the IH mass
region. 
The $\nu $-e cross section including CC and NC interactions 
 with atomic electrons is given as \cite{hoo71}
\begin{equation}
\frac{\sigma _e(E)}{dT}=\frac{2G_F^2m_e}{\pi}[g_L^2+g_R^2(1-\frac{T}{E_{\nu}})^2-g_Rg_L\frac{m_eT}{E_{\nu}}],
\end{equation}
where $E_{\nu}$ is the neutrino energy, $T$ is the electron kinetic energy, $G_F$ is the Fermi constant, $g_L=0.5 + sin^2\theta _W$, and $g_R =sin^2\theta _W$=0.231. 
Noting  that the 0$\nu \beta \beta $ $Q$-value for
the most relevant DBD
nuclei of current interest is around 2.5 $\pm$0.5 MeV, the only solar-$\nu$ component 
to be considered at the ROI is  the $^8$B-$\nu$ flux.
The recoil electron yield $b(E)$ as a function of the energy is derived from the cross section $\sigma _e(E)$ and the solar-$\nu$ flux $\phi(E)$ corrected for the $\nu $-oscillation effect \cite{bar11}. 

We consider a liquid scintillation detector with $N'$ 
tons of the scintillator and $N$ tons of the DBD 
isotope dissolved into the scintillator.  
In most DBD experiments using liquid scintillators, one may assume $N' \gg N$, and thus the ratio is $N/N' \ll$1. Furthermore we assume a constant detector volume. 
The neutrino-electron scattering rate at the energy window $\Delta E$ of ROI per year per ton (yt) of the DBD isotope is expressed as 
\begin{equation}
B_e(E)=0.6 \times b(E) \frac{Z'}{A'} \frac{E\delta }{R} ~/yt,
\end{equation}
where $b(E)$ is the electron yield in unit of 10$^{-33}$ per keV per year per electron in the liquid scintillator, and $Z'/A'$ is the
 ratio of the number of electrons to that of 
nucleons in the scintillator, $\delta =\Delta E/E$ stands for the energy resolution and $E=Q$ is the DBD $Q$ 
value in units of MeV. The energy window $\Delta E$ may correspond to the FWHM resolution in units of MeV.  
The electron to nucleon ratio is $Z'$/$A'$=50/92 in case of toluene (C$_6$H$_5$CH$_3$) and $Z'$/$A'$=130/232 for LAB
(assuming an average molecule of C$_{17}$H$_{28}$), respectively. For simplicity $Z'/A'$=0.55 is chosen.  

The energy spectrum of the recoiling electrons is nearly flat as a function of the energy \cite{bar11} in the region of 2-3 MeV. It is approximately given as 
$b(E)\approx 0.67-0.063E$ with $E$ being the scattered electron energy in units of MeV.
Thus the event rate is nearly independent of the DBD $Q$ values, and
the rate is given as 
\begin{equation}
B_e(E)\approx 0.15 \times E f ~/y t,
\end{equation}
where $f=\delta /R$ with $R=N/N'$ being the ratio (concentration) of the DBD isotopes to the scintillator ones. $f$  is a kind of a  BG efficiency for the liquid scintillator. As  the resolution $\delta $ (i.e. 
$\Delta E$) increases, $f$ and thus $B_e(E)$ increase. As the concentration of the DBD isotopes decreases, the relative weight of the DBD isotopes decreases and 
 $f$ and $B_e(E)$ relative to the DBD isotope increases. 
In a typical case of $\delta $=5$\%$ and $R$=1$\%$, one gets $f=\delta/R$=5 
and $B_e(E) \approx $ 2-3 /yt at $E\approx $3 MeV.

Next we consider the CC interaction on carbon nuclei in the scintillator. The cross section 
 is given as \cite{eji00a} 
\begin{equation}
\sigma _C(k) = \frac{G^2_F cos^2\theta _c}{\pi} p_e E_e F(Z,E_e) B_k ~~cm^2,
\end{equation}
where $G_F$ is the Fermi weak coupling constant, $\theta _c$ is the Cabibo angle, $p_e$ and $E_e$ are the out-going electron momentum and the total energy in MeV/c and MeV, $F(Z,E_e)$ is the Fermi-function
and $B_k$ is the weak strength for the $k^{th}$ state in the intermediate  nucleus B (see Fig.1). The strength is written as 
\begin{equation}
 B_k = B(F)_k + g^2_A B(GT)_k,
\end{equation}
where $B(F)_k$ and $B(GT)_k$ are the Fermi and GT strengths, 
and $g_A$ =1.267 is the axial vector coupling constant in unit of the vector one \cite{mun13}.
The CC cross section can be rewritten as
\begin{equation}
\sigma _C (k)= 1.6 \times 10^{-44} p_e E_e F(Z,E_e) B_k ~~cm^2.
\end{equation}
Since  $^{12}$C with $Q$ = -17.338 MeV cannot be excited by the solar-$\nu$ CC interaction, the CC interaction to be considered 
is the $^8$B-$\nu $ CC interaction on $^{13}$C with the natural abundance of 1.1$\%$.
The interaction is predominantly the ground state ($k=gs$) transition with $Q$ = -2.220 and log$ft$ = 3.7. 
We evaluate the BG rate  $B_{gs}(E)$  at $E=Q$=3 MeV of  current interest.  
Then the CC interactions of the $^8$B-$\nu$ with the energy around $E_{\nu}$=5.22 MeV contributes to the BG in the ROI  at $E$=3 MeV. 
Using the CC cross section for $E$=3 MeV and the $^8$B-$\nu$ flux at 5.22 MeV taking
$\nu$ oscillations into account, 
the BG rate integrated over the energy window of $\Delta E=E \delta$ is of the order of $10^{-3} f E$/yt. This is 
much smaller than the BG rate of the $\nu$-e  scattering (eq.(3)).\\
 
\section{Comparison with DBD rate}
We now evaluate the 0$\nu \beta \beta $ signal rate to compare with the solar-$\nu $ BG rates.
The 0$\nu \beta \beta $ rate per yt  for the light Majorana-$\nu $ exchange is written as  \cite{eji05,ver12}
\begin{equation}
S_{0\nu} = ln2 ~G^{0\nu}(m_{eff})^2 [M^{0\nu}]^2 \epsilon \frac{6 \times 10^{29}}{A}~ /yt,
\end{equation}
where  $G^{0\nu}$ is the phase space volume, $m_{eff}$ is the effective Majorana $\nu $-mass in unit of the electron mass , 
$\epsilon$ is the 0$\nu \beta \beta $ peak detection 
efficiency after various on-line and off-line cuts and  $M^{0\nu}$ is the nuclear matrix element for the light $\nu$-mass process.
Here we note that $G^{0\nu}$ includes conventionally the axial weak coupling $g_A=1.267g_V $ with $g_V$ being the vector coupling constant. Then   
$M^{0\nu}$ is expressed as 
\begin{equation}
M^{0\nu}=[\frac{g_A^{eff}}{g_A}]^2 M^{0\nu}_A({\rm NM})+[\frac{g_V^{eff}}{g_A}]^2 M^{0\nu}_V({\rm NM}),
\end{equation}
where $M^{0\nu}_A({\rm NM})$ and $M^{0\nu}_V({\rm NM})$ are axial-vector and vector components, and $g^{eff}_A$ and $g^{eff}_V$ are
the effective axial-vector and vector coupling constants in nuclei. The ratios $g^{eff}_A/ g_A$ and  $g^{eff}_V/ g_V$ stand for the renormalization(quenching) factors
 due to such non-nucleonic (isobar, exchange current etc) and nuclear medium effects that are not explicitly included in the nuclear matrix components \cite{eji78,eji09}.  Actually,
 the renormalization factor around  $g^{eff}_A/g_A\approx 0.5 -0.6 $ is suggested in case of the pnQRPA model due to isobar and other nuclear medium effects \cite{eji14A,eji15}). 

For a typical case of  $G^{0\nu}= 5 \times 10^{-14}$/y, $m_{eff}$ = 20 meV/$m_e$, $A$=100, and $\epsilon$=0.6,
 the 0$\nu \beta \beta $ signal rate is 
\begin{equation}
S_{0\nu}\approx 0.2 \times (M^{0\nu})^2  ~/ yt. 
\end{equation}
In the case of  a typical  $M^{0\nu}\approx$ 2-3, the signal rate is an order of 1/yt, 
This is of the same order of magnitude as the solar-$\nu$ BG rate for $f\approx 2$ and $E\approx 3$ MeV given in eq.(3).

The $0\nu \beta \beta $ signal rates for $^{100}$Mo and $^{130}$Te are plotted as a function of the effective $\nu $-mass and
 are compared with the $\nu $-e scattering BG rates in Fig.2. 
The DBD signal rate with $M^{0\nu}$=2 for the IH mass of 20 meV is nearly the same as the BG rate for a very ideal case
of $f$=1, for example this could be 5 $\%$ energy resolution and 5$\%$ DBD isotope concentration. 
In a more realistic case of $f$=5, i.e. 7$\%$ resolution and 1.4 $\%$ concentration,
the BG rate is larger than the signal rate at $m_{eff}$ =20 meV, and thus it is hard to cover the IH mass region of 15-45 meV.

\begin{figure}[htb]
\caption{The 0$\nu \beta \beta $ signal rates for $M^{0\nu}$=2, i.e. $S_{0\nu}$/$s^2$ with $s=M^{0\nu}/2$ , 
as a function of the effective $\nu $-mass  $m_{eff}$ and 
the solar-$\nu $ BG rates 
 for $^{100}$Mo, $^{130}$Te  and $^n$Te (thin line). Given are  the rates per year 
 per ton of  $^{100}$Mo, $^{130}$Te  and $^n$Te. 
The solar-$\nu$ e scattering BG rates  $B_l$ /y t  are shown for $f$=1 and 5 (horizontal lines).
\label{fig:strengths}}
\begin{center}
\includegraphics[width=0.6\textwidth]{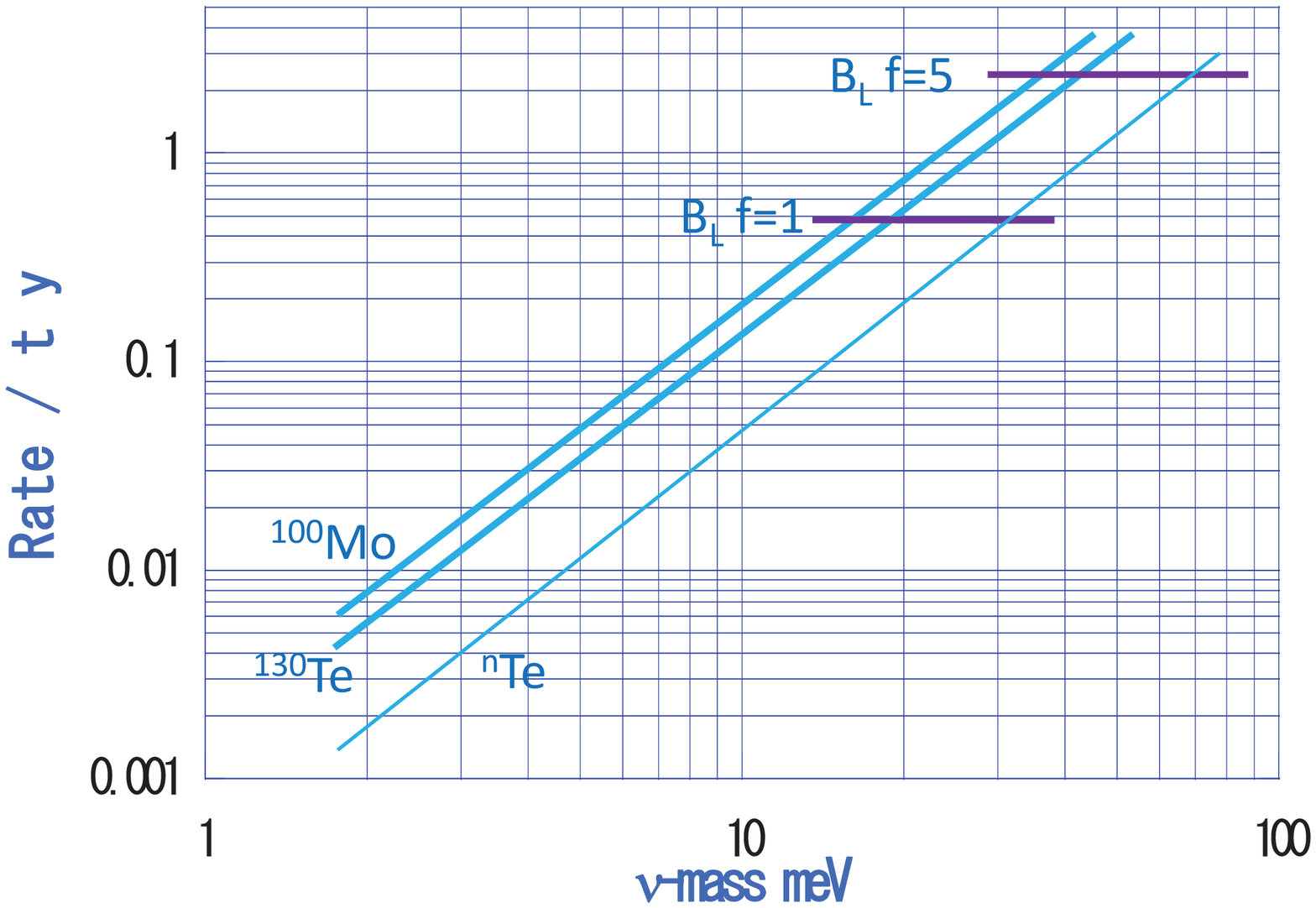}
\end{center}
\end{figure}

The $0\nu \beta \beta $ rate given in eq.(7) is the rate in case of pure (100$\%$ enriched) DBD isotopes. 
 If one uses natural Te ($^n$Te) isotopes with 34$\%$ $ ^{130}$Te, the $^{130}$Te $\beta \beta $ signal rate per ton of $^n$Te  gets 
smaller by a factor 3 than the rate for the enriched $^{130}$Te, while the solar-$\nu $ BG rate remains same, as shown in Fig.2. 
Accordingly, one needs a liquid scintillator with $f=\delta /R \ll$1 to cover the IH mass region if $M^{0\nu}$ is around or less than 2, 
while the requirement is a bit relaxed to $f=\delta /R \ll$2  if $M^{0\nu}$ is as large as 3.
 \\
\section{Neutrino mass sensitivity}
In order to discuss quantitatively the neutrino mass to be detected in DBD experiments, we introduce the neutrino mass  
sensitivity $m_{\nu}$ defined as the minimum Majorana $\nu $-mass to be detected by the detector with 90 $\%$ confidence level. 
The minimum mass in unit of MeV is written as \cite{eji05,ver12}
 \begin{equation}
 m_{\nu}=\frac{78}{M^{0\nu} G^{1/2} \epsilon ^{1/2}}\frac{n^{1/2} }{(NT)^{1/2}}~meV,
 \end{equation}
 where  $G=G^{0\nu}/(0.01 A)$ with $A$ being the mass number, 
  $\epsilon $ is the 0$\nu \beta \beta $ peak detection efficiency, as defined before,  $n$ is the number of counts required to identify the 0$\nu \beta \beta $ peak with 90 $\%$ confidence level, 
and $T$ is the exposure time in units of year.
  In case of ideal experiments with no BGs,  we get $n\approx $ 2.3. In practice, the BG rate is so large that one may set
 $n$ =1.7 $\times (BNT)^{1/2} $  with $B$ being the BG rate per year per ton of the DBD isotopes \cite{eji05}.

Then the mass sensitivity is rewritten as
\begin{equation}
 m_{\nu}=\frac{m^0 f^{1/4}}{(NT)^{1/4}},
 \end{equation}
 \begin{equation}
m^0= 102 \times \frac{(0.33~ b(E)E)^{1/4}}{M^{0\nu}G^{1/2}\epsilon ^{1/2}}~meV,
 \end{equation}
 where $m^0$ is the unit mass sensitivity in units of meV and $E$=$Q$ is in units of MeV. This is the sensitivity for $NT$=1 (ton year) and $f$=1, 
for example  $R$=10  $\%$ loading and $\delta $=10 $\%$ resolution.  The background rate $b(E)$ is obtained from the energy spectrum 
of the scattered electrons \cite{bar11}.

Extensive studies of high-sensitivity DBD experiments are in progress on
such DBD nuclei as $^{76}$Ge, $^{82}$Se, $^{100}$Mo, $^{116}$Cd, $^{130}$Te, and $^{136}$Xe \cite{eji05,ver12}. 
Here we assume the 0$\nu \beta \beta $ peak efficiency $\epsilon $=0.6 after various cuts.
The $m^0$ values together with the $Q$ and $G$ values for these DBD nuclei  are listed in Table 1.

\begin{table}[htpb]
\caption{DBD nuclei and the unit mass sensitivities $m^0$ in units of meV.
$Q$ is the 0$\nu \beta \beta $ $Q$ value in MeV, $G$ = $G^{0\nu }$/(0.01$A$), and 
$b(E)$ is the BG rate in units of 10$^{-33}$ per year per keV per electron of the liquid scintillator at $E$=$Q$.  
The column 5 shows $m^0/(2/M^{0\nu})$, i.e. the unit mass sensitivity for $M^{0\nu}$=2. 
\label{tab:masses1}}
\centering
\vspace{0.5cm}
\begin{tabular}{ccccc}
Isotope & $Q$ MeV & $G$ 10$^{-14}$/y  & $b(E) $ /keV/ ye & $m^0/(2/M^{0\nu})$    \\
\hline
$^{76}$Ge & 2.039 & 0.93 & 0.57 & 50 \\
$^{82}$Se & 2.998 & 3.79 & 0.51 & 29\\
$^{100}$Mo & 3.034 & 5.03 & 0.50& 24 \\
$^{116}$Cd & 2.814 & 4.69 & 0.52& 25\\
$^{130}$Te & 2.528 & 3.76 & 0.53 & 28\\
$^{136}$Xe & 2.468 & 3.77 & 0.53 & 28 \\
$^{150}$Nd & 3.368 & 15.5 & 0.53 & 15\\

\hline
\end{tabular}
\end{table}
The unit mass sensitivities  for DBD nuclei of current interests are plotted in Fig.3. 
The sensitivities are around 25-30 meV for most DBD nuclei 
except the cases of $^{76}$Ge and $^{150}$Nd, where the sensitivities are around 50 meV and 15 meV
respectively. 
The large $m^0$ value for $^{76}$Ge is due to the relatively small phase space of $G^{0\nu}$.
However, from the experimental point of view, $^{76}$Ge experiments have the best energy resolution. 

\begin{figure}[htb]
\caption{The unit mass sensitivities for DBD nuclei of current interests. The values $m^0/(2/M^{0\nu})$ (dark blue squares), 
i.e. the unit mass sensitivities $m^0$  
in case of the matrix element $M^{0\nu}$=2 and the 100 $\%$ enriched isotope. The values $m^0/r^{1/2}$ (light blue squares) are for 
natural Te with $r$=34$\%$ of $^{130}$Te and natural Nd with $r$=5.6$\%$ of $^{150}$Nd respectively (see text).
\label{fig:unit}}
\begin{center}
\includegraphics[width=0.55\textwidth]{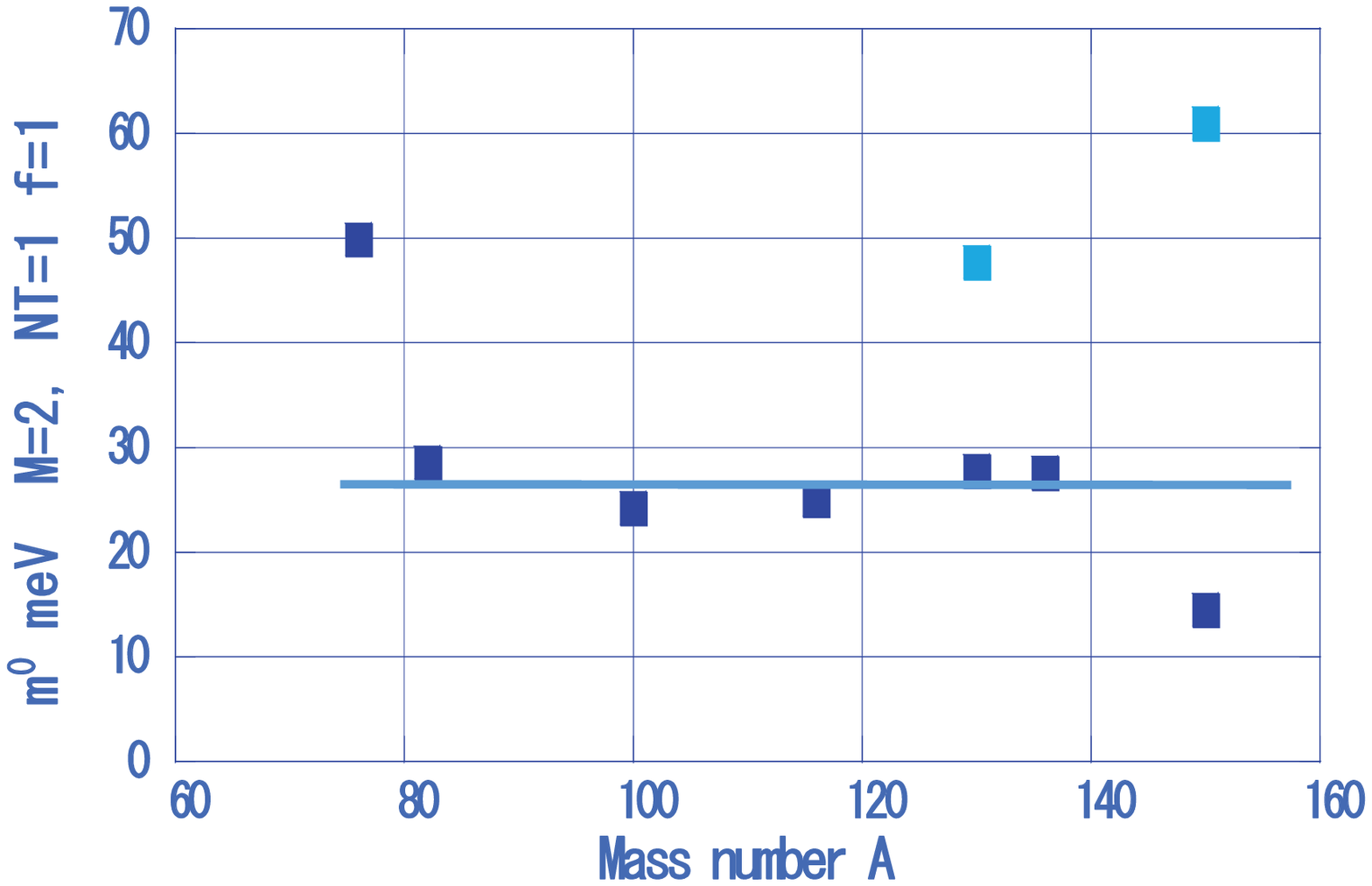}
\end{center}
\end{figure}

As an example, the mass sensitivities for $^{130}$Te are plotted 
for three cases of the liquid scintillators with $f$ = 1,  5 and 20 in Fig.4. 
 One needs around 50 ton year exposures in cases of $f$=5 to cover the IH mass region of 15-45 meV. 
The mass sensitivities for other isotopes
 are nearly the same except $^{76}$Ge and $^{150}$Nd since the unit mass-sensitivities are nearly the same as shown in Fig.3.
 
\begin{figure}[htb]
\caption{Neutrino mass sensitivities $m_{\nu}$ with $M^{0\nu}$ =2  for liquid scintillation detectors with $^{130}$Te 
for three cases of $f$=1, 5 and 20.
\label{fig:strengths}}
\begin{center}
\includegraphics[width=0.6\textwidth]{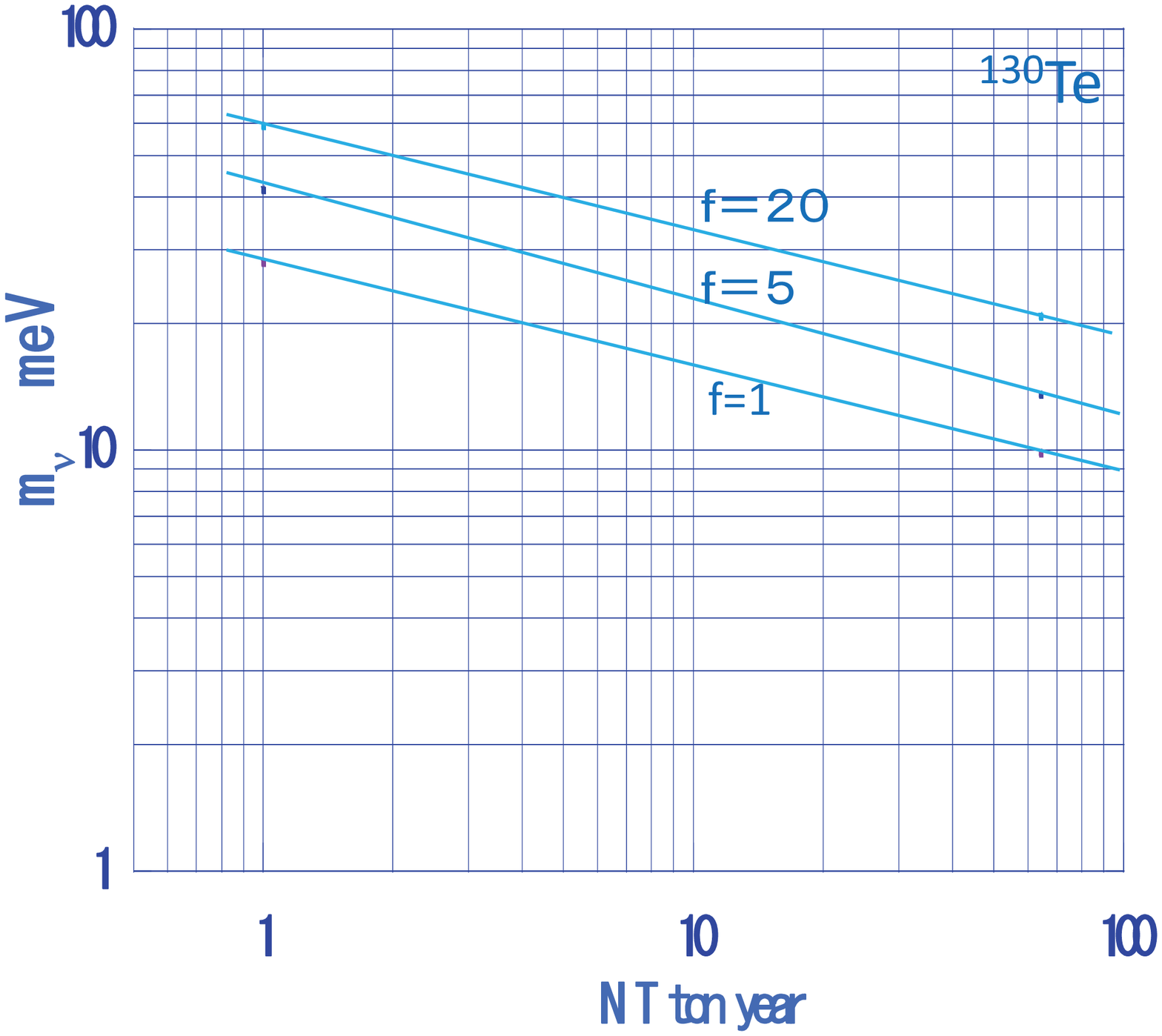}
\end{center}
\end{figure}

\begin{figure}[htb]
\caption{Neutrino mass sensitivities $m_{\nu}$ with $M^{0\nu}$ =2 for $f$=5 liquid scintillators with $^{130}$Te and $^n$Te are plotted
 as function of the $NT$ yt of $^{130}$Te and $^n$Te, respectively.
\label{fig:strengths}}
\begin{center}
\includegraphics[width=0.6\textwidth]{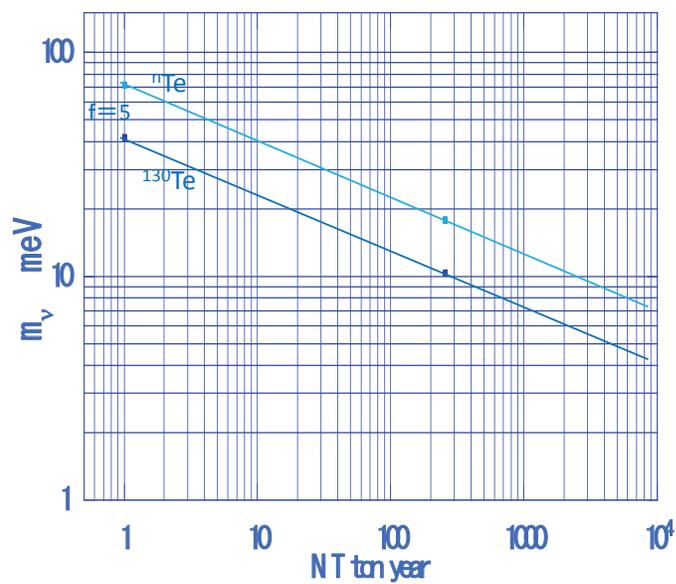}
\end{center}
\end{figure}

\begin{figure}[htb]
\begin{center}
\end{center}
\end{figure}

The mass sensitivity given in eq.(10) is for the case of 100 $\%$ enriched DBD isotope. In case of DBD isotopes with enrichment $r$, the 
0$\nu \beta \beta $ peak efficiency $\epsilon$ gets effectively reduced by the factor $r$ and the
unit sensitivity is given by $m^0r^{-1/2}$.  
 DBD isotopes with $r \approx$0.8-0.9 are used if the centrifugal isotope enrichment is realistic.  Then the mass sensitivities are 10-5$\%$ larger than the 
values for the 100$\%$ enrichment.  
The mass sensitivity for the $N$ ton natural Te ($^n$Te) with  $^{130}$Te abundance of $r$=0.34 is worse  
 than that for pure $^{130}$Te by a factor 
$r^{-1/2}$=1.7 , as shown in Figs. 4 and 5. 
In order to achieve the same sensitivity as
 the enriched $^{130}$Te, one needs an order of magnitude more isotopes if the exposure time is fixed. \\
 
\section{Summary}

 Solar neutrinos interactions with atomic electrons in liquid scintillation detectors used for DBD experiments are  evaluated. 
 They are summarized as  follows.
 
 1. The BG rate at ROI(region of interest for the 0$\nu \beta \beta $ experiment) is given by 
$B_e(E)\approx 0.15 \times E f$ /y t (year ton),
where $f$ is the ratio of the detector energy resolution and the concentration of the DBD isotopes in the scintillator. It is
of the same order of magnitude as the 0$\nu \beta \beta $ signal rate for the IH $\nu $-mass. So they are serious BG sources 
for DBD experiments to study the IH $\nu$-mass region. 

2. Solar-$\nu $ interactions are not avoidable, but their BG contributions are reduced by 
improving the resolution $\delta $ and/or
increasing the concentration $R$. In case of $M^{0\nu}\approx $2, scintillation detectors with $f\approx $1 
may cover the IH mass of 15-45 meV by the exposure of $NT\approx $10 ty, but those with $f\approx $5 require $NT\approx $50 ty.
Then detectors required to
 cover the NH mass region of $m_{\nu }$ = 1.5 - 4 meV  would be 4 orders of magnitude larger in scale than those for the IH mass 
region, if $f$ would remain the same.
 
3. It is important to use enriched isotopes, if possible, 
to increase the signal rate and  to decrease the BG rate per ton of the DBD isotopes. In other words,
experiments with only a few $\%$ abundance of a DBD isotope suffer 
from large contributions of the solar-${\nu }$ interactions with atomic electrons and nuclei in other isotopes. Thus 
they should be enriched to more than 30-50\% to cover even the IH $\nu$ mass.  In case of  $^{48}$Ca 
with the natural abundance  of 0.2 $\%$, enrichment to around 50 $\%$ is crucial although the BG rate $b(E)$ 
at $Q$=4.3 is lower by 15$\%$ than those for other isotopes with $Q\approx $ 3 MeV \cite{bar11}.

4. So far we discussed mainly the BG rates from the liquid scintillator with a small
 concentration of $R\ll$1. If $R$ is increased to a couple of 10 $\%$ 
in order to improve the mass sensitivity (eq.10), it may affect the resolution $\delta $ and increase 
 the BG rate from DBD atomic electrons and nuclei, too. The density change in case of heavy loading may require modifications of the support system of the scintillation 
apparatus. For example in SNO+ with low loading, ropes have to be fixed at the ground floor, while for massive loading they have
to be fixed on the top as the density becomes higher than the surrounding water. Similar issues might be occurring at other experiments as well. Otherwise it could only be compensated by removing scintillator if the volume is fixed. 
Furthermore it would be a big challenge to reduce the
 resolution to a few $\%$ and to increase the concentration to an order of 10 $\%$ simultaneously to realize the small BG 
efficiency of $f \ll$1  as required to cover the full IH mass region.  
 
5. The scattered electron is a forward-peak distribution with respect to the incident solar-$\nu$ direction. 
As scintillation detectors also contain a weak contribution of Cerenkov light which is directional sensitive and very fast
it might be used, in principle, for rejecting the solar-$\nu$ events, but whether it can be 
done has to be shown.

\section{Remarks}

 Brief remarks on solar neutrino contributions to DBD experiments are given in this section. The BG contributions due to the solar-$\nu $ interactions with DBD nuclei should also be well considered in case of liquid scintillators with modest energy resolution 
of $\delta $. If the BG events involve $\beta$,$ \gamma $, and/or  inverse $\beta $ rays as shown in Fig.1, they might be reduced by
 measuring their spacial correlations (Signal Selection by Spacial Correlation SSSC) 
and/or their time correlations (Signal Selection by Time Correlation SSTC) \cite{eji05,eji14}. 
If not, gamma rays from excited states will add to the background as well.

 It should be noted that the mass sensitivity does depend on $M^{0\nu}$. The sensitivities for $M^{0\nu} $=2 as plotted in Figs. 4 and 5  are smaller or larger by a factor 1.3
 if the matrix element gets larger or smaller by the same factor. In fact,  $M^{0\nu}$ is rather model dependent, and 
subjects to the uncertainties of the interaction parameters and the renormalization (quenching) of the weak coupling $g_A$ used 
in the model \cite{ver12,eji78,eji09, eji14A,eji15,suh14,bar13}.  \\

In the present letter, we have discussed general features of the solar-$\nu$ BG contributions 
in liquid scintillator DBD experiments, and have evaluated them. The solar-$\nu$ ES BGs were also discussed in a review \cite{ell04}. 
 Actually, they depend 
on the energy resolution $\delta $ (which might change with loading due to
changes in optical properties) and the DBD isotope concentration $R$, and also on the way to 
reduce the BGs by various hard-ware and analysis cuts. Consequently, the effects of the solar-$\nu$ interactions 
depend largely on details of the individual detector specifications, the configurations and the data-analyses. So we do not 
discuss the actual rates of the solar-$\nu$ BGs for individual current and future detectors.   

Finally, we remark that the solar-$\nu$ interactions might be potential BG sources for any large-scale experiments to search for low-energy rare events as DBD and dark matter \cite{ver08,bar14}, 
or in turn these detectors could be used for solar neutrino studies as well if $\beta \beta $ and other BGs would be eliminated  or separated well.\\\

The authors thank  Dr. B. von Krosigk for valuable discussions.

\end{document}